\documentclass[superscriptaddress,nofootinbib,twocolumn]{revtex4}
\usepackage{amsmath,lscape,epsfig}
\usepackage{amsfonts}
\usepackage{amsmath}
\usepackage{amssymb}
\usepackage{graphicx}

\def\ii{\'{\i}}
\def\beq{\begin{equation}}
\def\eeq{\end{equation}}
\def\beqa{\begin{eqnarray}}
\def\eeqa{\end{eqnarray}}
\def\ban{\begin{eqnarray*}}
\def\ean{\end{eqnarray*}}
\def\bi{\begin{itemize}}
\def\ei{\end{itemize}}

\begin{document}

\title{Quark matter under strong magnetic fields in the su(3) 
Nambu--Jona-Lasinio Model}

\author{D.P. Menezes}
\affiliation{Depto de F\'{\i}sica - CFM - Universidade Federal de Santa
Catarina  Florian\'opolis - SC - CP. 476 - CEP 88.040 - 900 - Brazil}
\author{M. Benghi Pinto}
\affiliation{Depto de F\'{\i}sica - CFM - Universidade Federal de Santa
Catarina  Florian\'opolis - SC - CP. 476 - CEP 88.040 - 900 - Brazil}
\author{S.S. Avancini}
\affiliation{Depto de F\'{\i}sica - CFM - Universidade Federal de Santa
Catarina  Florian\'opolis - SC - CP. 476 - CEP 88.040 - 900 - Brazil}
\author{C. Provid\^encia}
\affiliation{Centro de F\ii sica Computacional - Department of Physics -
University of Coimbra - P-3004 - 516 - Coimbra - Portugal}

\begin{abstract}

In the present work we use the mean field  approximation to investigate quark 
matter described by the su(3)  Nambu--Jona-Lasinio model subject to a strong 
magnetic field. We consider two cases: pure quark matter and quark matter 
in $\beta$-equilibrium possibly present in magnetars. The results are 
compared with the ones obtained with the su(2) version of the model.
{ The energy per baryon of magnetized quark matter becomes more bound 
than nuclear
matter made of iron nuclei, for $B$ around $2 \times 10^{19}$ G.
When the $su(3)$ NJL model is applied to stellar matter}, the maximum mass 
configurations are always above 1.45 $M_\odot$ and may be
as high as 1.86 $M_\odot$ for a central magnetic field of $5\times 10^{18}$ G.
These numbers are within the masses of observed neutron stars.  
\end{abstract}

\maketitle

\vspace{0.50cm}
PACS number(s): {24.10.Jv,26.60+c,11.10.-z,11.30Qc}
\vspace{0.50cm}

\section{Introduction}

{ In non-central heavy ion collisions such as
the ones performed at RHIC and LHC-CERN, physicists have been looking for  
a possible signature of the presence of CP-odd domains in the presumably
formed quark-gluon plasma phase \cite {qgp}.
The study of deconfined quark matter subject to strong external magnetic 
fields is then mandatory if one intends to understand the physics taking place 
in such colliders.

Neutron stars with very strong magnetic fields of the order of
$10^{14}-10^{15}$ G are known as magnetars and they are believed to be the
sources of the intense gamma and X rays detected in 1979 \cite{duncan,kouve}.
The hypothesis that some neutron stars are constituted by unbound quark matter 
cannot be completely ruled out \cite{quarks} since the Bodmer-Witten conjecture
\cite{conjecture} cannot be tested on earthly experiments. This conjecture 
implies that the true ground state of all matter is (unbound) quark matter 
because theoretical predictions show that its energy per baryon at zero 
pressure is lower than $^{56}$Fe binding energy.
   
In the present work our aim is to investigate quark matter described by the 
su(3) version of the Nambu-Jona-Lasinio \cite{njl} model exposed to strong 
magnetic fields. In the case of pure quark matter, as predicted by the QCD 
phase transition possibly taking place in heavy ion collisions, the magnetic 
field is certainly external. In the case of neutron stars, the magnetic field 
can be generated by the alignment of charged particles that are spinning very
rapidly. We 
next use an external field to mimic the real situation, which we do not know 
how to determine. Albeit in an approximate way, the effect of the 
magnetic field on the macroscopic quantities as radius and masses can be 
obtained.

Recently the su(2) version of the NJL model was used to treat both situations 
described above \cite{prc}. We have shown that, for pure quark matter,
the energy per baryon for magnetized quark matter has a minimum which is lower 
than the one determined for magnetic free quark matter. 
We have also obtained that a magnetic field of the order
of $2 \times 10^{18}\,\,$G barely affects the effective mass as compared 
with the results for matter not subject to the magnetic field. For 
$B=5\times 10^{19}$ G matter is totally polarized for chemical potentials 
below 490 MeV. For small values of the magnetic fields the number of filled 
Landau levels (LL) is large and the quantisation effects are
washed out, while for large magnetic fields the chiral symmetry restoration
occurs for smaller values of the chemical potentials.
When $\beta$-equilibrium is enforced, the numerical results show that, for the 
the su(2) case, only very high magnetic fields ($B \ge 10^{18} \,$G) affect 
the equation of state (EOS) in  a noticeable way.

The inclusion of the $s$-quarks, necessary in the su(3) NJL model, poses some 
new numerical difficulties and some questions that need to be addressed. Those
problems are tackled through out the paper. One of the questions was raised
in \cite{buballa96,buballa99} and refers to the stability of quark matter 
described by the NJL model. The authors show that it is not absolute stable.
As already mentioned, in \cite{prc} we have seen that the inclusion of the 
magnetic field increases stability in the su(2) version and the same behavior is
expected in the su(3) NJL, which is shown next.

The paper is organized is such a way that all calculations already shown 
explicitly in \cite{prc} are not repeated but all important differences are 
outlined. In sections II and III the formalism (mean field theory) and the 
equations of state are shown and in section IV the final results are displayed 
and the conclusions are drawn.}

\section{General formalism}

In order to consider (three flavor) quark stars in $\beta$ equilibrium with
strong magnetic fields
one may  define the following lagrangian density
\begin{equation}
{\cal L} = {\cal L}_{f}+{\cal L}_{l} - \frac {1}{4}F_{\mu
\nu}F^{\mu \nu}
\end{equation}
where the quark sector is described by the  su(3) version of the  Nambu--Jona-Lasinio model 
\begin{equation}
{\cal L}_f = {\bar{\psi}}_f \left[\gamma_\mu\left(i\partial^{\mu}
- q_f A^{\mu} \right)-
{\hat m}_c \right ] \psi_f ~+~ {\cal L}_{sym}~+~{\cal L}_{det}~,
\label{njl}
\end{equation}
where ${\cal L}_{sym}$ and ${\cal L}_{det}$ are given by:
\begin{equation}
{\cal L}_{sym}~=~ G \sum_{a=0}^8 \left [({\bar \psi}_f \lambda_ a \psi_f)^2 + ({\bar \psi}_f i\gamma_5 \lambda_a
 \psi_f)^2 \right ]  ~,
\label{lsym}
\end{equation}
\begin{equation}
{\cal L}_{det}~=~-K \left \{ {\rm det}_f \left [ {\bar \psi}_f(1+\gamma_5) \psi_f \right] + 
 {\rm det}_f \left [ {\bar \psi}_f(1-\gamma_5) \psi_f \right] \right \} ~,
\label{ldet}
\end{equation}
where $\psi_f = (u,d,s)^T$ represents a quark field with three flavors, ${\hat m}_c= {\rm diag}_f (m_u,m_d,m_s)$ is the corresponding (current) mass matrix while $q_f$
represents the quark electric charge and $\lambda_a$ denotes the Gell-Mann matrices. Here, we consider $m_u=m_d \ne m_s$. The  ${\cal L}_{det}$ term is the t'Hooft interaction which represents a determinant in flavor space which, for three flavor, gives a six-point interaction \cite {buballa} 
\begin{equation} 
 {\rm det}_f ({\bar \psi}_f {\cal O} \psi_f):=\sum_{i,j,k} \epsilon_{ijk} ({\bar u} {\cal O} \psi_i)
({\bar d} {\cal O} \psi_j)({\bar s} {\cal O} \psi_k) \,\, ,
\end{equation}
{ where $\epsilon_{ijk}$ is the usual three-dimensional Levi-Civita symbol.
The lagrangian also contains the ${\cal L}_{sym}$ term which is symmetric
under global  $U(N_f)_L\times U(N_f)_R$  transformations and corresponds to a
4-point interaction in flavor space. In the appendix we discuss the steps to
obtain ${\cal L}_f$ in the mean-field approximation (MFA).}

The leptonic sector is given by
\begin{equation}
\mathcal{L}_l=\bar \psi_l\left[\gamma_\mu\left(i\partial^{\mu} - q_l A^{\mu}
\right) -m_l\right]\psi_l \,\,,
\label{lage}
\end{equation}
where $l=e,\mu$. One recognizes this sector as being represented by
the usual QED type of lagrangian density. As  usual, $A_\mu$ and $F_{\mu \nu }=\partial
_{\mu }A_{\nu }-\partial _{\nu }A_{\mu }$ are used to account
for the external magnetic field. Then, since we are interested in a
  static and constant magnetic field
in the $z$ direction, $A_\mu=\delta_{\mu 2} x_1 B$.
\section {The EOS}
We need to evaluate the thermodynamical potential for the three flavor quark sector, $\Omega_f$, which as usual can be written as $\Omega_f = -P_f = {\cal E}_f - T {\cal S} - \sum_f \mu_f \rho_f $ where $P_f$
represents the pressure, ${\cal E}_f$ the energy density, $T$ the temperature,
${\cal S}$ the entropy density, and
$\mu_f$ the chemical potential. 

For the present study, just the zero temperature
case is important and, as a consequence, the term with the entropy vanishes. The total pressure for three flavor in $\beta$ equilibrium is given by
\begin{equation}
P(\mu_f,\mu_l,B)= P_f^N |_{M_f}+ P_l^N |_{m_l} +\frac{B^2}{2}
\,\,,
\end{equation}
where our notation means that $P_f^N$ is evaluated in terms of  the quark effective mass, $M_f$, which is
determined in a (nonperturbative) self consistent way while $P_l^N$ is evaluated at the leptonic bare mass, $m_l$. The term  $B^2/2$ arises due to the electromagnetic  term $F_{\mu \nu}F^{\mu \nu}/4$ in the
original lagrangian density. The subscript $N$ indicates normalized pressures. Here, our normalization choice is such that $P_f^N=0$ at $\mu_f=0$ ($f=u,s,d$) and $P_l^N=0$ at $\mu_l=0$ ($l=e,\mu$) implying that $P(0,0,B)=B^2/2$.
\subsection{Quark Contribution to the EOS}
In the mean field approximation the pressure can be written as
\begin{equation}
P_f = \theta_u+\theta_d+\theta_s 
-2G(\phi_u^2+\phi_d^2+\phi_s^2) + 4K \phi_u \phi_d \phi_s \,\,,
\end{equation}
where an irrelevant term has been discarded. The pressure due to the three quarks is diagrammatically represented in figure 1a. 
\begin{figure}[h]
\begin{tabular}{ccc}
\end{tabular}
\end{figure}
\begin{figure}[h]
\begin{tabular}{ccc}
\includegraphics[width=7.cm]{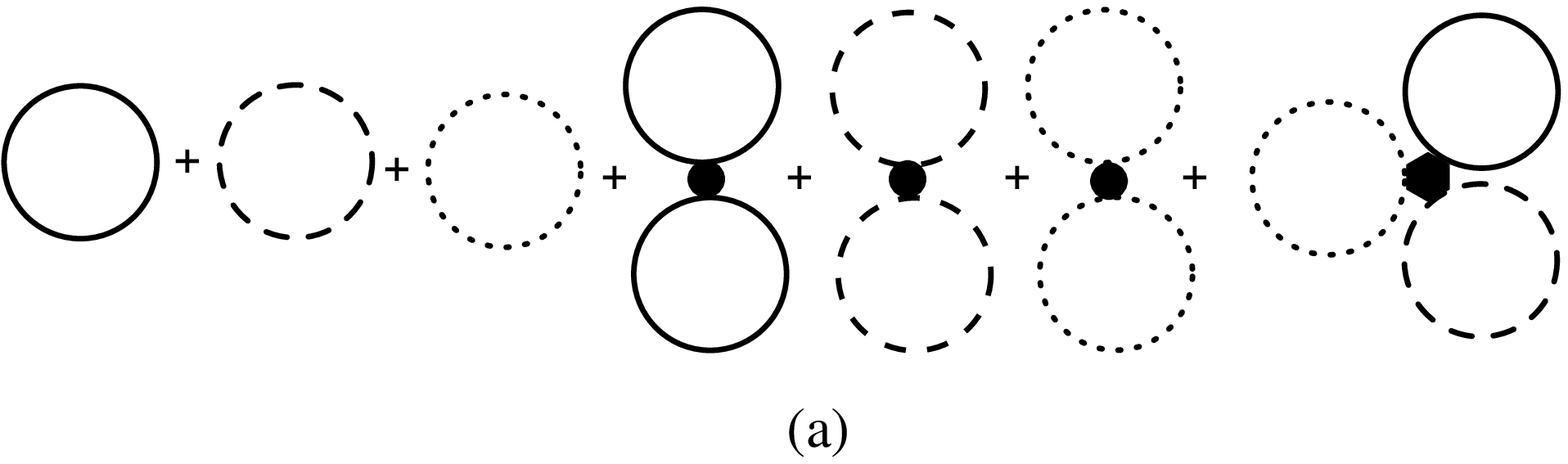}\\
\includegraphics[width=7.cm]{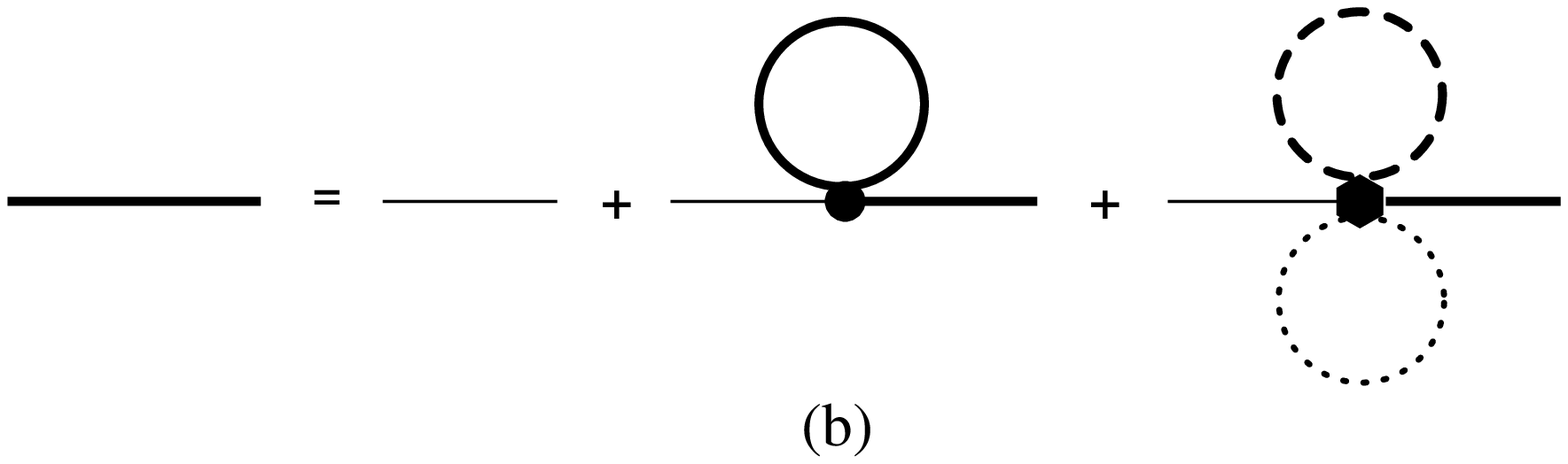}  
\end{tabular}
\caption{a) Feynman diagrams contributing to the quark pressure in the MFA. The lines represent the three dressed quark propagators for the different flavors: $u$ (continuous line), $d$ (dashed line) and $s$ (dotted line). The black dot represents $G$ and the black hexagon represents $K$. b) Diagrammatic representation of the effective mass for flavor $u$. The diagrams contributing to the other two flavor display the same topology.}
\label{figsnovas}
\end{figure}

For a given flavor, the  $\theta_f$ term is given by
\begin{equation}
 \theta_f=-\frac{i}{2}  {\rm tr}  \int  \frac {d^4 p}{(2\pi)^4} \ln \left(-p^2 + M_f^2 \right )
\end{equation}
and the condensates, $\phi_f$ are given by
\begin{equation}
\phi_f= \langle {\bar \psi}_f \psi_f \rangle= -i  \int \frac {d^4 p}{(2\pi)^4} {\rm tr}\frac{1}{(\not \!
p - M_f+i\epsilon)}
\label{cond}
\end{equation}
where all the traces are to be taken over color ($N_c=3$) and Dirac space, but not flavor.  
In order to obtain results valid at finite $T$ and $\mu$ in the presence of an external magnetic field $B$
one can use the following replacements

\[
p_0 \to i(\omega_\nu - i \mu_f)\,\,\,,
\]
\[
{\bf p}^2 \to p_z^2 +(2n+1-s) \;\;\;\;,{\rm with}\;\;\;\;\;s=\pm 1\,\, \;\;,\;n=0,1,\dots
\]
\[
\int \frac{d^4 p}{(2\pi)^4} \to i \frac{T |q_f| B}{2\pi} \sum_{\nu = -\infty}^{\infty} \sum_{n=0}^{\infty} \int \frac{d p_z}{(2\pi)} \,\,\,\,.
\]
In the above relations, $\omega_\nu= (2 \nu+1)\pi T$, with
$\nu=0,\pm 1,\pm 2,\ldots$ representing the Matsubara frequencies
for fermions while $n$ represents the Landau levels (LL) and $s$
represents the spin states which, at $B \ne 0$, must be treated
separately. The case $T=0$ in which we are interested can be easily obtained after the above 
substituions (see Ref. \cite{prc}).

The effective quark masses can be obtained self consistently  from (see figure 1b)
\begin{equation}
 M_i=m_i - 4 G \phi_i + 2K \phi_j \phi_k, 
 \label{mas}
\end{equation}
with $(i,j,k)$ being any permutation of $(u,d,s)$. So, to determine the EOS for the su(3) NJL at finite density and in the presence of a magnetic field  we need to know the condensates, $\phi_f$, as well as the contribution from the gas of quasiparticles, $\theta_f$. Both quantities, which are related by $\phi_f \sim d \theta_f  / dM_f$, have been evaluated with great detail in Ref. \cite{prc}. Here, we just quote the results
 \begin{equation}
P_f= \left (P^{vac}_f+P^{mag}_f + P^{med}_f \right )_{M_f}\,\,,
\label{pressBmu2}
\end{equation}
where the vacuum contribution reads

\begin{equation}
P^{vac}_{f}=- \frac{N_c }{8\pi^2} \left \{ M_f^4 \ln \left [
    \frac{(\Lambda+ \epsilon_\Lambda)}{M_f} \right ]
 - \epsilon_\Lambda \, \Lambda\left(\Lambda^2 +  \epsilon_\Lambda^2 \right ) \right \},
\end{equation}
where we have defined $\epsilon_\Lambda=\sqrt{\Lambda^2 + M_f^2}$ with $\Lambda$ representing a non covariant ultra violet cut off. The evaluations performed in Ref. \cite{prc} also give the following finite  magnetic contribution
\begin{equation}
P^{mag}_f= \frac {N_c (|q_f| B)^2}{2 \pi^2} \left [ \zeta^\prime(-1,x_f) -  \frac {1}{2}( x_f^2 - x_f) \ln x_f +\frac {x_f^2}{4} \right ]\,\,,
\end{equation}
where   $x_f = M_f^2/(2 |q_f| B)$ while
$\zeta^\prime(-1,x_f)= d \zeta(z,x_f)/dz|_{z=-1}$ where $\zeta(z,x_f)$ is the Riemann-Hurwitz zeta function \cite {wolfram}. Finally, after integration, the medium contribution can be written as
\begin{eqnarray}
P^{med}_{M_f}&=&\sum_{k=0}^{k_{f,max}} \alpha_k\frac {|q_f| B N_c }{4 \pi^2}  \left [ \mu_f \sqrt{\mu_f^2 - s_f(k,B)^2} \right .\nonumber \\
&-& \left . s_f(k,B)^2 \ln \left ( \frac { \mu_f +\sqrt{\mu_f^2 -
s_f(k,B)^2}} {s_f(k,B)} \right ) \right ] ,
\label{PmuB}
\end{eqnarray}
where  $s_f(k,B)
= \sqrt {M_f^2 + 2 |q_f| B k}$, $\alpha_0=1,\,\alpha_{k>0}=2$.
The  upper Landau level (or the nearest integer) is defined by
\begin{equation}
k_{f, max} = \frac {\mu_f^2 -M_f^2}{2 |q_f|B}= \frac{p_{f,F}^2}{2|q_f|B}.
\label{landaulevels}
\end{equation}

Finally, the condensates $\phi_f$ entering the quark pressure  at finite density and in the presence of an external magnetic field can also be written as 

\begin{equation}
\phi_f=(\phi_f^{vac}+\phi_f^{mag}+\phi_f^{med})_{M_f}
\end{equation}
where

\begin{eqnarray}
\phi_f^{vac} &=& -\frac{ M_f N_c }{2\pi^2} \left [
\Lambda \epsilon_\Lambda -
 {M_f^2}
\ln \left ( \frac{\Lambda+ \epsilon_\Lambda}{{M_f }} \right ) \right ]\,\,,
\end{eqnarray}

\begin{eqnarray}
\phi_f^{mag}
&=& -\frac{ M_f |q_f| B N_c }{2\pi^2}\left [ \ln \Gamma(x_f)  \right . \nonumber \\
&-& \left .\frac {1}{2} \ln (2\pi) +x_f -\frac{1}{2} \left ( 2 x_f-1 \right )\ln (x_f) \right ] \,\,,
\end{eqnarray}
and
\begin{eqnarray}
\phi_f^{med}&=&
\sum_{k=0}^{k_{f,max}} \alpha_k \frac{ M_f |q_f| B N_c }{2 \pi^2} \nonumber \\
&\times& \left [\ln \left ( \frac { \mu_ f +\sqrt{\mu_f^2 -
s_f(k,B)^2}} {s_f(k,B)} \right ) \right ]\,\,.
\label{MmuB}
\end{eqnarray}
From the pressure one can obtain the density, $\rho_f$, corresponding to each different flavor, which is given by

\begin{equation}
\rho_f=
\sum_{k=0}^{k_{f,max}} \alpha_k \frac{|q_f| B N_c }{2 \pi^2} k_{F,f} \,\,,
\end{equation}
where  $k_{F,f}=\sqrt{\mu_f^2 - s_f(k,B)^2}$, since $dP/d\phi_f=0$.

The quark contribution to the energy density is

\begin{equation}
{\cal E}_{f} (\mu_f,B)= -P_{f}^N + \sum_{f} \mu_f \rho_f \,\,\,,
\end{equation}
 where $P_f^N = P_f(\mu_f)|_{M_f(\mu_f)} - P_f(0)|_{M_f(0)}$.

Throughout this paper we consider the following set of parameters
\cite {buballa}:  $\Lambda = 631.4 \, {\rm MeV}$ , $m_u= m_d=\,  5.5 {\rm MeV}$,
$m_s=\,  105.66 {\rm MeV}$, $G \Lambda^2=1.835$ and $K \Lambda^5=9.29$.

\subsection{Lepton Contribution to the EOS}

The leptonic contribution, $P_l$ has also been evaluated in detail in Ref. \cite {prc} where the normalization requirement $P_l^N=0$ at $\mu_l=0$  has  been adopted.
The result  shows that, at the one loop level, only the following (finite) medium contribution has to be considered
\begin{eqnarray}
P_{l}^N&=&\sum_{l=e}^\mu \sum_{k=0}^{k_{l,max}} \alpha_k\frac {|q_l| B }{4 \pi^2}   \left [ \mu_l \sqrt{\mu_l^2 - s_l(k,B)^2} \right .\nonumber \\
&-& \left . s_l(k,B)^2 \ln \left ( \frac { \mu_l +\sqrt{\mu_l^2 - s_l(k,B)^2}}
    {s_l(k,B)}   \right ) \right ]~.\nonumber \\
&&
\end{eqnarray}
Then, the leptonic density is also easily evaluated yielding
\begin{equation}
\rho_l =  \sum_{k=0}^{k_{l, max}}\alpha_k \frac{ |q_l| B }{2 \pi^2}   k_{F,l}(k,s_l) \,\,,
\label{rholmuB}
\end{equation}
where $k_{F,l}(k,s_l) =\sqrt{\mu_l^2 - s_l(k,B)^2}$.
Finally, the leptonic energy density reads
\begin{equation}
{\cal E}_{l} (\mu_l,B)= -P_{l}^N + \sum_{l} \mu_l \rho_l \,\,\,.
\end{equation}

The lepton masses are
$m_e=0.511 \, {\rm MeV}$ and  $m_\mu=105.66\, {\rm MeV}$.

\section{Results and conclusions}

In the sequel we consider two different situations of quark matter
  under a strong magnetic field: a)  pure
  quark  matter with the same  chemical potential for all quark flavors; b)
  $\beta$-equilibrium quark stellar matter.

\begin{figure}[thb]
\includegraphics[width=0.8\linewidth,angle=0]{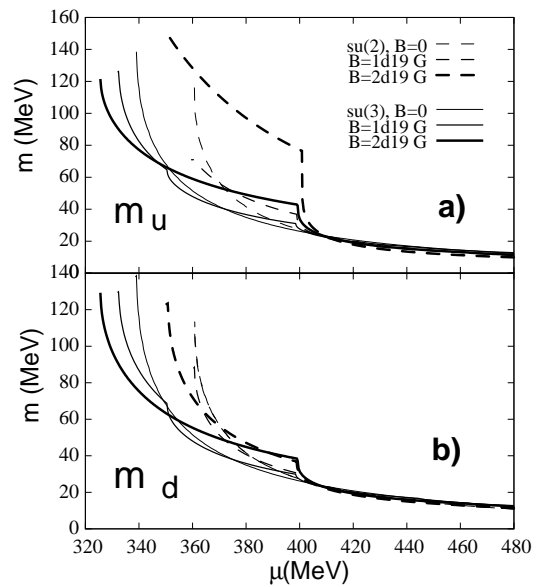}
\caption{Mass of the quarks (a) $u$ and (b) $d$ as a function of the chemical potential  
for $B=0,\, 10^{19}$ and $2\times 10^{19}$ G within the su(2) NJL and su(3) 
NJL.}
\label{mud}
\end{figure}

We first discuss the properties of pure quark matter with equal
  chemical potentials for all flavors, namely the behavior of the dynamical
quark  masses, the chiral symmetry restoration with density and  the energy
per baryon.
In Fig. \ref{mud} we display the masses of quarks $u$ and $d$ as function
of the chemical potential for different values of the magnetic field and the two
versions of the NJL model.  For the magnetic field intensities used,  one can clearly identify the filling of different 
Landau levels 
causing the usual kinks in the curves. For the three intensities considered the chiral symmetry is approximately restored for $\mu=400$ MeV.

It is interesting to see that although 
the general behavior is the same, the effect of the LL is more pronounced in the
su(2) version.   

\begin{figure}[thb]
\includegraphics[width=0.8\linewidth,angle=0]{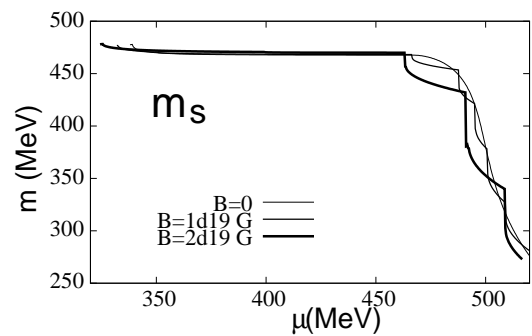}
\caption{Mass of the $s$ quarks as a function of the chemical potential for $B=0,\, 10^{19}$ 
and $2\times 10^{19}$ G within $su(3)$ NJL.} 
\label{ms} 
\end{figure}

In Fig. \ref{ms} the mass of the $s$ quark is shown as a function
of the chemical potential for different values of the magnetic field. 
One can see how drastically it falls  around $\mu=450$ MeV. For magnetic free 
quark matter, this is the same behavior shown in Fig. 3 of \cite{schaffner}. 
One can observe that the curve is no longer smooth when $B$ is turned on, but
the values of the strange quark mass do not vary much. According to 
\cite{schaffner}, the fact that the strange quark mass remains relatively high
as compared with the masses of the other two quarks is the main reason why
deconfined quark matter may not be likely to appear in the core of hybrid 
neutron stars. For a magnetic field larger than 10$^{19}$ the restoration of
chiral symmetry for the $s$-quark occurs in steps and starts at a smaller chemical potential than the B=0 case. 

\begin{figure}[thb]
\includegraphics[width=0.8\linewidth,angle=0]{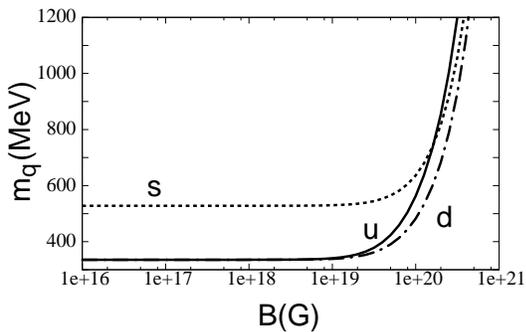}
\caption{Vacuum mass of the quarks as a function of the magnetic field B.}
\label{vacuum}
\end{figure}

The phenomenon of magnetic catalysis, which enhances chiral symmetry breaking, has been well discussed within the $su(2)$ version of the NJL model \cite {klimenko}. Here, 
for reference, we  show the vacuum effective mass of the three quarks as a function of the magnetic field in Fig. \ref{vacuum}. For $B>10^{19}$ G the vacuum masses increase dramatically with the magnetic field as expected. A similar  increase of the vacuum mass was also obtained for the $su(2)$ version of NJL in \cite{klimenko,prc} and the effect is related to the fact that the B field facilitates the binding
by antialigning the helicities of the quark and the antiquark,
which are then bound by the NJL interaction. As shown in Fig. \ref{vacuum}, an interesting result of the $su(3)$ version is that, due to its larger  electric
charge, the $u$ quark has an effective mass that becomes larger than that   
of the $s$ quark for $B>1.5\times 10^{20}$ G.

\begin{figure}[thb]
\includegraphics[width=0.8\linewidth,angle=0]{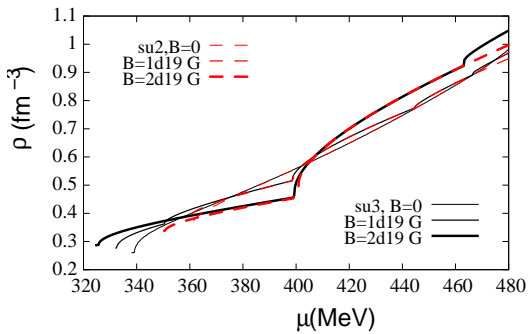}
\caption{Baryonic density as a function of the quark chemical potential for 
$B=0,\, 10^{19}$ and $2\times 10^{19}$ G within both $su(2)$ and $su(3)$ NJL.} 
\label{mr}
\end{figure}

In Fig.\ref{mr} the baryonic density is shown as a function of the 
quark chemical potential for two values of the magnetic field and  for
both versions of the NJL model, $su(2)$ and $su(3)$. As already noticed in 
\cite{prc}, once again, 
for small values of the magnetic fields the number of filled LL is quite 
large and the effects of the quantization are less visible.
Due to the Landau quantization, the increase of the strength of the
magnetic field gives rise to a decrease of the number of the filled
LL and the amplitude of the oscillations is more clear in the
graphics. For each value of the magnetic field, the kink
appearing at the smallest chemical potential corresponds to the case when
only the first LL has been occupied.

\begin{figure}[thb]
\includegraphics[width=0.8\linewidth,angle=0]{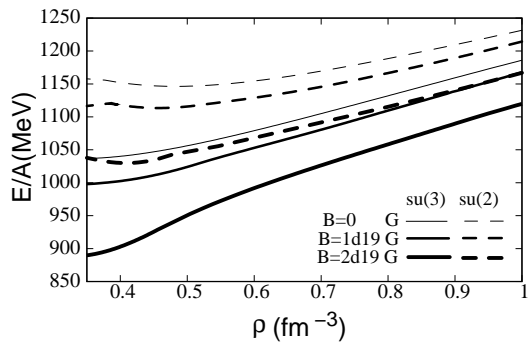}
\caption{Energy per nucleon  as a function of density for $B=0,\, 10^{19}$ 
and $2\times 10^{19}$ G within NJL $su(2)$ and NJL $su(3)$.}
\label{ebin}
\end{figure}

In Fig. \ref{ebin} one can see that the inclusion of the
magnetic field makes matter more and more bound in both versions of the model.
For the present set of parameters, the energy per baryon $E/A$ 
of magnetized quark matter { becomes more bound than nuclear
matter made of iron nuclei, $\frac{E}{A}|_{^{56}Fe} \sim 930$ MeV
for $B$ around $2 \times 10^{19}$ G.}

\begin{figure}[thb]
\includegraphics[width=0.8\linewidth,angle=0]{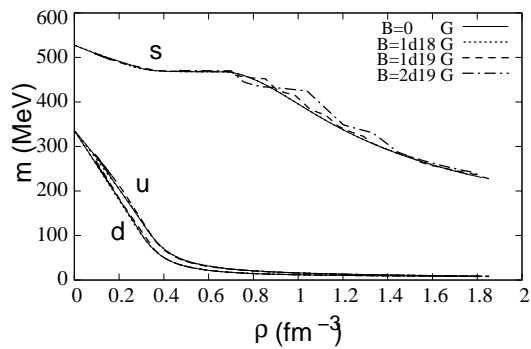}
\caption{The quark effective mass for $\beta$-equilibrium quark matter with a 
constant magnetic field within  NJL $su(3)$.}  
\label{meffsu3a}
\end{figure}

\begin{figure}[thb]
\includegraphics[width=0.8\linewidth,angle=0]{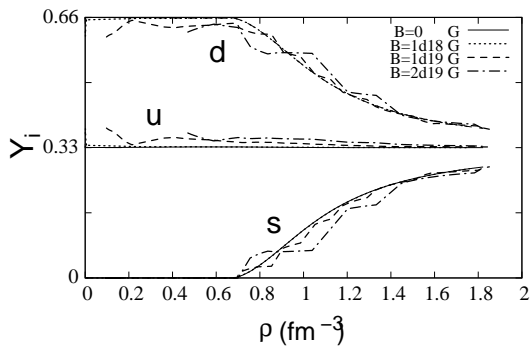}
\caption{Fraction of quarks in $\beta$-equilibrium quark matter for a 
constant magnetic field within  NJL $su(3)$.}  
\label{fraction_su3a}
\end{figure}

We next consider stellar matter made out of quarks, electrons and muons 
in $\beta$-equilibrium, as possibly occurring  
in the interior of magnetars. It is worth mentioning that, in this case, the 
three different quarks bear different chemical potentials, determined by the chemical equilibrium conditions
$$\mu_d=\mu_s=\mu_u+\mu_e, \quad \mu_\mu=\mu_e.$$
We start by plotting the quark effective masses 
for different values of the magnetic field in Fig. \ref{meffsu3a}. 
It is seen that the results for non-magnetized matter ($B=0$) almost coincide 
with  the ones obtained for $B=10^{18}$G. A decrease of the $s$ quark mass
starts only at $\sim 0.8$ fm$^{-3}$. This behavior had already been discussed
in \cite{dp04}.
If the magnetic field is strong enough the mass of quark $s$ occurs in finite jumps which may give rise to an increase of the strangeness fraction as shown in  Fig. \ref{fraction_su3a}.

The quark fractions $Y_i=\rho_i/\rho$, $i=u,d,s$ are shown in Fig. 
\ref{fraction_su3a}. Again the results for $B=0$ are similar to the ones for
  $B=10^{18}$G. For strong enough fields the quark $u$ fractions increase with a reduction of the quark $d$ fraction. The quark $s$ fraction has a sudden increase for $\rho\sim 0.7$ fm$^{-3}$ but above   $\rho\sim 0.9$ fm$^{-3}$ remains below the $B=0$ fraction.

\begin{figure}[thb]
\includegraphics[width=0.8\linewidth,angle=0]{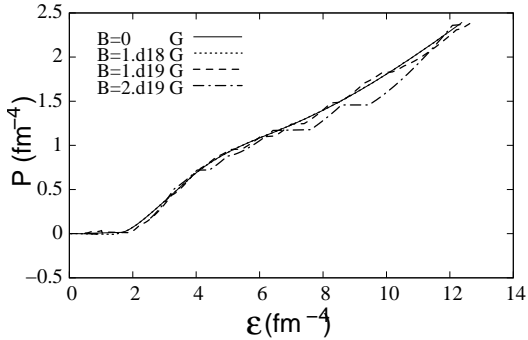}
\caption{Equation of state of $\beta$-equilibrium quark matter for constant 
magnetic fields within  the NJL $su(3)$.} 
\label{eos}
\end{figure}

In Fig. \ref{eos} the EOS for different values of the magnetic field is 
shown. For magnetic fields as large as  $B=10^{18}$G the differences are very 
small as compared with non-magnetized matter. For larger fields there is an overall net softening of the EOS.

\begin{figure}[thb]
\includegraphics[width=0.8\linewidth,angle=0]{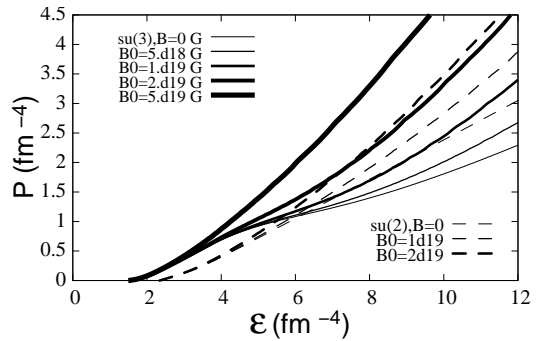}
\caption{Equation of state of $\beta$-equilibrium quark matter for a density 
dependent magnetic field within NJL $su(2)$ and NJL $su(3)$. 
The EOS for $B=0$ is also shown.}
\label{eosb}
\end{figure}

\begin{figure}[thb]
\includegraphics[width=0.8\linewidth,angle=0]{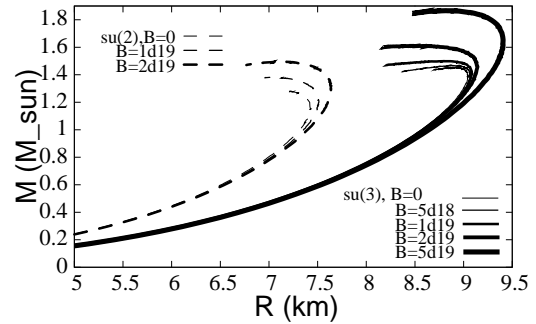}
\caption{Mass-radius curves for the  families of stars within NJL $su(2)$ and NJL $su(3)$ corresponding to the EOS shown in Fig. \ref{eos}. For $B\le 10^{18}$ the curves coincide with the $B=0$ results.}
\label{tov}
\end{figure}

It is well known that at the surface the 
magnetic field should not be larger than $\sim 10^{15}$ G. We have introduced 
a density dependent magnetic field as in \cite{chakra97,prc}:
\beq
B_i =B^{\hbox{surf}} + B_0\left[1-\exp\left\lbrace-\beta\left( \frac{\rho_b}{\rho_0}\right)^\gamma  \right\rbrace  \right],
\label{bvar}
\eeq 
where $B^{\hbox{surf}}=10^{15}$ G is  the magnetic
 field  at the surface, $B_i$ is the magnetic field at the interior of the 
star for large densities and the 
parameters $\alpha=5\times 10^{-5}$ and $\gamma=3$ were chosen in such a way 
that the field increases fast with density to its central value but still  
describes correctly the surface of the star where the pressure is zero. We show
the equations of state for quark matter in $\beta$-equilibrium and a density 
dependent magnetic field within both versions of the NJL model in 
Fig. \ref{eosb}. As implicit in
eq. (\ref{bvar}), the field at the surface is $10^{15}$ G. The magnetic field 
makes the EOS harder  with consequences in the gravitational 
and baryonic masses of compact stars, whose properties are obtained from the 
integration of the
Tolman-Oppenheimer-Volkoff equations, which use as input the EOS obtained with 
the density dependent magnetic field. The results are displayed both in 
Fig. \ref{tov} and in Table \ref{tabela}, from where it is seen that both the
gravitational and the baryonic masses increase with the increase of the 
magnetic field for an intensity larger than  $\sim 5 \times 10^{18}$ G for the 
su(3) version and $10^{18}$ G for the su(2) NJL. However, the increase of the
gravitational mass is larger than the increase of the
baryonic mass because the contribution of the magnetic field becomes more  and
more important as the field increases. This  explains the decrease of the
central energy/baryonic  density for the stronger fields considered.

 Another
important effect of the field on the properties of the stars is the increase
{ of the radius of the star with the largest radius}, which may be as high as 9.5 Km for the $su(3)$
NJL. In general, the maximum mass star configurations for the  $su(2)$ version of
the NJL model  are smaller with smaller radius, $\sim 7$ Km, in average 2 Km
smaller than the corresponding stars in the $su(3)$ version of
the NJL model.  

Within the $su(3)$
NJL the maximum mass configurations are always above 1.45 $M_\odot$ and may be
as high as 1.86 $M_\odot$ for a central magnetic field of $5\times 10^{18}$ G.
These numbers are within the masses of observed neutron stars. On the other
hand the  $su(2)$ version of the NJL model forsees too small star masses  
except for very large magnetic fields. 

\begin{table}
\caption{Maximum mass configurations for  NJL $su(3)$ and $su(2)$, and  several magnetic field intensities: the gravitational mass ($M$), baryonic mass ($M_b$), radius ($R$), central energy density ($\epsilon_c$), baryonic density ($\rho_c$) and magnetic field ($B_c$) are given}
\begin{tabular}{lcccccc}
\hline
$B_0$& $M$ & $M_b$& $R$ & $\epsilon_c$ &  $\rho_c$&$B_c$\\
(G)& ($M_\odot$) &($M_\odot$) &(Km) & fm$^{-4}$ & fm$^{-3}$ &(G)\\
\hline 
$su(3)$\\
0                 &  1.46 &1.53& 8.93&7.49 &1.19 & 10$^{15}$\\
         10$^{18}$&  1.46  &1.53& 8.93&7.49 &1.19 & 1.6$\times$10$^{17}$\\
5$\times$10$^{18}$&  1.47  &1.54& 8.88&7.94 &1.24 & 8.8$\times$10$^{17}$\\
1$\times$10$^{19}$&  1.50  &1.58& 8.78&8.36 &1.25 & 1.8$\times$10$^{18}$\\
2$\times$10$^{19}$&  1.61  &1.69& 8.53&9.64 &1.25 & 3.6$\times$10$^{18}$\\
5$\times$10$^{19}$&  1.86  &1.88& 8.81&9.26 & 1.01& 5.0$\times$10$^{18}$\\
$su(2)$\\
0&  1.29  &1.24& 7.09 &13.68 &1.86 &10$^{15}$\\
1$\times$10$^{18}$&  1.29  &1.25& 7.08&13.85 &1.88 & 1.2$\times$10$^{17}$\\
1$\times$10$^{19}$&  1.38  &1.33& 7.01&14.52 & 1.72 & 4.0$\times$10$^{18}$\\
2$\times$10$^{19}$&  1.49  &1.41& 7.11&14.47 &1.49 & 5.7$\times$10$^{18}$\\
\hline
\end{tabular}
\label{tabela}
\end{table}

The effects of the anomalous magnetic moments has been shown to be
relevant \cite{prakash,wei,aurora} and we intend to take them into account 
in the next calculations. 

The {  color superconductivity (CS)}
\cite{cfl},
which allows the quarks near the Fermi surface to form Cooper pairs
that condense and break the color gauge symmetry \cite{mga}
is known to be present in the QCD phase diagram at
sufficiently high densities. The effect  of strong magnetic fields on the CS properties
of quark matter, which can be drastic for sufficiently high fields, has
already been studied by several authors \cite{b-cs}. It would be important to
investigate how this SC phase could affect the properties of quark stars under
strong magnetic fields. However, it could be that CS is only affected by magnetic fields
stronger than the ones considered in the present paper, which, however,
predicts already a very high maximum mass, $M\sim 1.9$ M$_\odot$. The largest magnetic
field we got in the center of a quark star is $5\times 10^{18}$ G, while in
\cite{b-cs} it is shown that a noticeable effect requires fields above
$\sim 10^{19}$ G.

\section*{Acknowledgments}

This work was partially supported by the Capes/FCT n. 232/09 bilateral collaboration,
 by CNPq (Brazil),  by FCT and FEDER (Portugal) under the project
CERN/FP/83505/2008 and  by Compstar, an ESF Research
Networking Programme.

%
%
%

\appendix

\section{The su(3) NJL model in the MFA}
In this appendix the main steps in order to obtain the Nambu-Jona-Lasinio
lagrangian given eq.(\ref{njl}) in the mean field approximation { are
explicitly shown.} Firstly, we consider the ${\cal L}_{sym}$ term given in 
eq.(\ref{lsym}). For later convenience, we define the matrix elements of 
$\Phi$ and its adjoint $\Phi^{\dagger}$ as \cite{hatsuda}:
\begin{equation}
\Phi_{ij} = \bar{\psi}_j(1-\gamma_5)\psi_i ~,~
\Phi^{\dagger}_{ij} = \bar{\psi}_j(1+\gamma_5)\psi_i  \nonumber ~,
\label{phi}
\end{equation}
where i, j are flavor labels.  From these definitions, one can easily show that:
\begin{eqnarray}
\bar{\psi}_f(1-\gamma_5)\lambda_a \psi_f~&=&~tr(\lambda_a \Phi)~,~ \\ \nonumber
\bar{\psi}_f(1+\gamma_5)\lambda_a \psi_f~&=&~tr(\lambda_a \Phi^{\dagger}) ~,
\end{eqnarray}
where $tr$ is the trace operator in flavor space. So, adding and subtracting these expressions, we can rewrite the NJL symmetric four-point interaction term   as:
\begin{eqnarray}
&& {\cal L}_{sym} ~=~ G \sum_{a=0}^8 \left [({\bar \psi}_f \lambda_ a \psi_f)^2 + ({\bar \psi}_f i\gamma_5 \lambda_a
 \psi_f)^2 \right ]  \nonumber \\
 &~=~& G \sum_{a=0}^8 ~tr(\lambda_a \Phi)~tr(\lambda_a \Phi^{\dagger})~= 2G ~tr( \Phi~ \Phi^{\dagger})~.
\label{eqlsym}
\end{eqnarray} 
The summation involved in the latter equality can be performed noting that an
arbitrary matrix $A$ in the $N_f$=3 flavor space, can be expanded in terms of 
Gell-Mann matrices as follows:
\begin{equation} 
 A=\sum_{a=0}^8 c_a \lambda_a ~,{\rm with} ~c_a=\frac{1}{2} tr(\lambda_a A)~.
\end{equation}
The expansion coefficients $c_a$  are obtained using the the Gell-Mann matrices property: $tr(\lambda_a \lambda_b)=2\delta_{ab}$. So, we can write:
\begin{equation}
tr(AA^{\dagger})=tr(\sum_{a=0}^8 c_a \lambda_a \sum_{b=0}^8 c^{\star}_b \lambda_b^{\dagger})~=~
\frac{1}{2}\sum_{a=0}^8 tr(\lambda_a A)tr(\lambda_a A^{\dagger}) ~,
\end{equation}
where in the latter term we have used that the Gell-Mann matrices are 
hermitian, i.e., $\lambda_a$ =$\lambda_a^{\dagger}$.
We then evaluate ${\cal L}_{sym}$ in the mean field approximation linearizing 
the interaction terms. We follow refs.\cite{buballa,hatsuda}  approximating 
the product of two operators $\hat{O}_1$ and $\hat{O}_2$ by:
\begin{eqnarray} 
\hat{O}_1 \hat{O}_2 \approx \hat{O}_1  \langle \hat{O}_2 \rangle ~+~
\langle \hat{O}_1  \rangle\hat{O}_2 ~-~
\langle \hat{O}_1 \rangle \langle \hat{O}_2 \rangle ~. \label{prod2}
\end{eqnarray}   
Therefore, calculating explicitly the trace involved in eq.(\ref{eqlsym}) and taking into account the prescription above,  ${\cal L}_{sym}$, can be written in the MFA as:
\begin{equation}
 {\cal L}_{sym} = 4G\left[ \phi_u ~u^{\dagger} u + \phi_d ~d^{\dagger} d
+ \phi_s ~s^{\dagger} s -\frac{1}{2} (\phi_u^2 + \phi_d^2 + \phi_s^2) \right]  ~, 
\label{lsmfa}
\end{equation} 
where we have used:
\begin{equation}
\langle \bar{\psi_i} \psi_j\rangle ~=~ \delta_{ij}\phi_i ~~{\rm and}~~ \langle \bar{\psi_i} \gamma_5 \psi_j\rangle ~=~0 ~. \label{condp}
\end{equation}
The only 3 non-vanishing terms are the condensates which were defined in eq.(\ref{cond}). 
Finally, we consider the t'Hooft term, eq.(\ref{ldet}),  which is a six-point interaction in the su(3) flavor space. Notice that term involves the product of three operators which we linearize analogously to eq.(\ref{prod2}):
\begin{eqnarray} 
&& \hat{O}_1 \hat{O}_2 \hat{O}_3  \approx   
\hat{O}_1  \langle \hat{O}_2  \rangle \langle \hat{O}_3  \rangle~+~
\langle \hat{O}_1  \rangle \hat{O}_2 \langle \hat{O}_3  \rangle ~+~
\langle \hat{O}_1  \rangle \langle \hat{O}_2  \rangle \hat{O}_3 \nonumber \\
&&~-~ 2 \langle \hat{O}_1 \rangle \langle \hat{O}_2 \rangle \langle \hat{O}_3  \rangle~. \nonumber
\end{eqnarray}
So, in the MFA the determinants  which appear in the t'Hooft term can be written as:

\begin{eqnarray} 
&&{\rm det}_f ({\bar \psi}_f {\cal O} \psi_f) = \sum_{i,j,k} \epsilon_{ijk} ({\bar u} {\cal O} \psi_i)
({\bar d} {\cal O} \psi_j)({\bar s} {\cal O} \psi_k) \approx \nonumber \\ 
&& \sum_{i,j,k} \epsilon_{ijk} [
({\bar u} {\cal O} \psi_i) \langle {\bar d} {\cal O} \psi_j \rangle \langle {\bar s} {\cal O} \psi_k \rangle ~+~
\langle {\bar u} {\cal O} \psi_i\rangle ({\bar d} {\cal O} \psi_j)\langle {\bar s} {\cal O} \psi_k\rangle \nonumber \\
&& ~+~ \langle {\bar u} {\cal O} \psi_i\rangle \langle {\bar d} {\cal O} \psi_j \rangle ({\bar s} {\cal O} \psi_k ) ~-~
2\langle {\bar u} {\cal O} \psi_i\rangle \langle{\bar d} {\cal O} \psi_j\rangle \langle {\bar s} {\cal O} \psi_k\rangle  \nonumber
\,\, ~.
\end{eqnarray}
Now, inserting the operator ${\cal O}=1\pm \gamma_5$ and using the properties given in eq.(\ref{condp}), we obtain the t'Hooft term in the MFA  :
\begin{equation}
{\cal L}_{det}= -2K ( \phi_d \phi_s \bar{u}u ~+~\phi_u \phi_s \bar{d}d ~+~\phi_u \phi_d \bar{s}s -
2 \phi_u \phi_d \phi_s ) ~.
\label{ldmfa}
\end{equation}
 From eq.(\ref{njl}) and eqs.(\ref{lsmfa},\ref{ldmfa}) the su(3) NJL lagrangian in the MFA is given by:
\begin{eqnarray}
{\cal L}_f^{MFA}  &=& {\bar{\psi}}_f \left( \gamma_\mu\left(i\partial^{\mu}
- q_f A^{\mu} \right)-
{\hat M} \right) \psi_f \nonumber \\
&& -2G (\phi_u^2 + \phi_d^2 + \phi_s^2) + 4K \phi_u \phi_d \phi_s ~,
\nonumber
\end{eqnarray}
where $\hat{M}$ is a diagonal matrix with  elements defined in eq.({\ref{mas}).

\end{document}